\documentclass[global,twocolumn,11pt]{llncs}

\usepackage{url}
\usepackage{graphicx}
\usepackage{listings}
\usepackage{semantic}
\usepackage{array}
\usepackage{a4wide}

\def\rapps{{\it rapps}}
\def\rapp{{\it rapp} }

\begin{document}

\title{Model Driven Reactive Applications}

\author{Tony Clark\inst{1} \and Dean Kramer\inst{2} \and Samia Oussena\inst{2}}

\institute{Middlesex University, London, UK, \email{t.n.clark@mdx.ac.uk}\and
University of West London, UK, \email{dean.kramer@uwl.ac.uk},\email{samia.oussena@uwl.ac.uk}}

\maketitle

\begin{abstract}
Reactive applications (\rapps) are of interest because
of the explosion of mobile, tablet and web-based platforms. The complexity
and proliferation of implementation technologies makes it attractive
to use model-driven techniques to develop \rapp systems. This article proposes a
domain specific language for \rapps\ consisting of
stereotyped class models for the structure of the application and state
machine models for the application behaviour. The models are given a semantics 
in terms of a transformation to a calculus called Widget. The 
languages are introduced using an example application for mobile phones.
\end{abstract}
\section{Introduction}

Harel and Pneuli \cite{harel1985development} define \emph{reactive
applications} (\rapps) as systems that receive events from their environment
and must react accordingly. Reactive systems are
of increasing interest partly because of the recent explosion in number and diversity of
mobile and tablet platforms. Together with web-applications, mobile
and tablet apps operate by reacting to user input and changes to the
platform context.

There are some characteristic features to this family of applications:
they are mainly driven by events that originate from the user or the application context; many applications have user-interfaces that consist of simple hierarchically organized elements such as text, buttons, input fields etc.; often applications can be described in terms of a machine whose states are described in terms of a tree of user-interface elements and associated event handlers, and whose transitions occur in response to events. Although the applications are essentially quite simple, the development complexity arises because of the significant differences between multiple target platforms.

\subsection{Model Driven Development}

Model Driven Development (MDD) \cite{france2007model} is an approach
to Software Engineering that uses models to abstract away from implementation
details and to use code generation or model execution to produce a
complete or partial system. By abstracting away from the implementation
technology, the system definition can target different 
platforms and it is argued that the system becomes easier to maintain \cite{DBLP:conf/icse/HutchinsonWRK11}.
Model Driven Architecture (MDA) is an approach that uses UML to perform
MDD and involves UML being used to construct Platform Independent
Models (PIMs) and Platform Specific Models (PSMs) and to model transformations
between them. MDD is of interest to \rapp development because of the
diversity and complexity of target platforms; an application can be
developed as a single model and then transformed to multiple implementation
technologies using general purpose transformations.

Although there are characteristic \rapp features, implementation technologies
remain general purpose. Android and iPhone applications are developed using 
frameworks where application classes extend platform specific libraries that 
hide the application logic. MDD approaches seek to address these issues by
abstracting away from the implementation details; however, current MDD approaches
that are relevant to \rapp are often incomplete and do not support reasoning about
the application.

This article describes work that aims to provide a precisely defined framework for
model-driven \rapp in terms of a modelling language and an associated calculus. Like other model-driven approaches, the structure and outline
behaviour of an application is specified using class diagrams and state machines 
(equivalent to other approaches that use class diagrams and activity diagrams).
However, we argue that approaches based purely on UML-style models, even with
action languages, lack the expressiveness necessary to capture application patterns and complex behaviour  such as call-backs. Such approaches are often based on 
stereotypes and lack analysis tools such as type-checkers. 
Therefore, we propose a calculus, called Widget, used to represent complete \rapp applications. Widget has a precisely defined operational
semantics and a type system that can be statically checked. Structure and behaviour 
diagrams are views of partial Widget programs and we define a translation 
from models to Widget.

\subsection{Problem and Contribution}

\label{problem}

The complexity and diversity of \rapp implementation platforms can be addressed
by suitable MDD approaches. However  current approaches are
lacking in terms of implementation independence, behavioural completeness, and support
for application analysis. A lack of  behavioural completeness compromises the model 
driven aims of these technologies in terms of being technology independent
since the code that is produced must be edited in order to run on
each target platform. Where there are many different target platforms
for a single application, this can be a significant task.

This article addresses the following problems in applying MDD techniques to \rapp development. Firstly, MDD techniques often use a {\it domain specific language} (DSL) to represent a family of related applications. There are candidate DSLs for \rapp development, however as described in section \ref{related_work} there are limitations in terms of completeness or consistency with \rapp implementation platforms. We perform a domain analysis that leads to a list of key features that must be supported by any DSL. Secondly, there is no generally accepted mechanism for expressing \rapp models. The Unified Modelling Language (UML) is the most widely used modelling notation in both academia and industry. Although UML supports features for general application development, and therefore can support \rapps, it is usual to support DSLs in UML via stereotypes. A stereotype is a specialization of a standard UML element that is tagged for a specific purpose, for example tagging a class as a relational database table. There is no set of stereotypes (or {\it profile}) for expressing \rapp models in UML and we use the domain analysis to derive a \rapp profile. Finally, detailed execution in UML models can be expressed using a general purpose action language that provides features similar to a standard programming language. Since the UML action language is general purpose it does not constitute a DSL for \rapp and therefore does not provide specific help for the verification of \rapp models. We present a calculus called Widget that is used as the action language for the \rapp profile. 

Widget is based on a functional language because it is simple and
universal. Functional languages are increasingly used as an alternative
traditional languages for web applications \cite{krishnamurthi2007implementation,serrano2006hop,cooper2007links}
partly because of the need for interactive applications to deal with
continuations and partly because of the interest in state-less concurrent
applications \cite{lammel2008google}. In addition the characteristic
features of \rapp applications are identified by adding them to a $\lambda$-calculus
in a simple way, for example using higher-order functions as event
handlers, continuations and to structure hierarchically organized
application objects. We use an approach based on monads to contain
those parts of an application that deal with updating state (SQLite
for example). As described in \cite{Peyton-Jones} this supports the
desirable situation where applications can be built from composable
units.

The languages are exemplified in terms of a context
aware application defined in \cite{daniele2009mda} called \emph{Buddy}.
The DSML is used to express the structure and state-transition behaviour 
of Buddy which is then translated to Widget. A Widget interpreter has
been developed in Java and used to implement the case study.

\begin{figure}[t]
\hfill{}\includegraphics[scale=0.35]{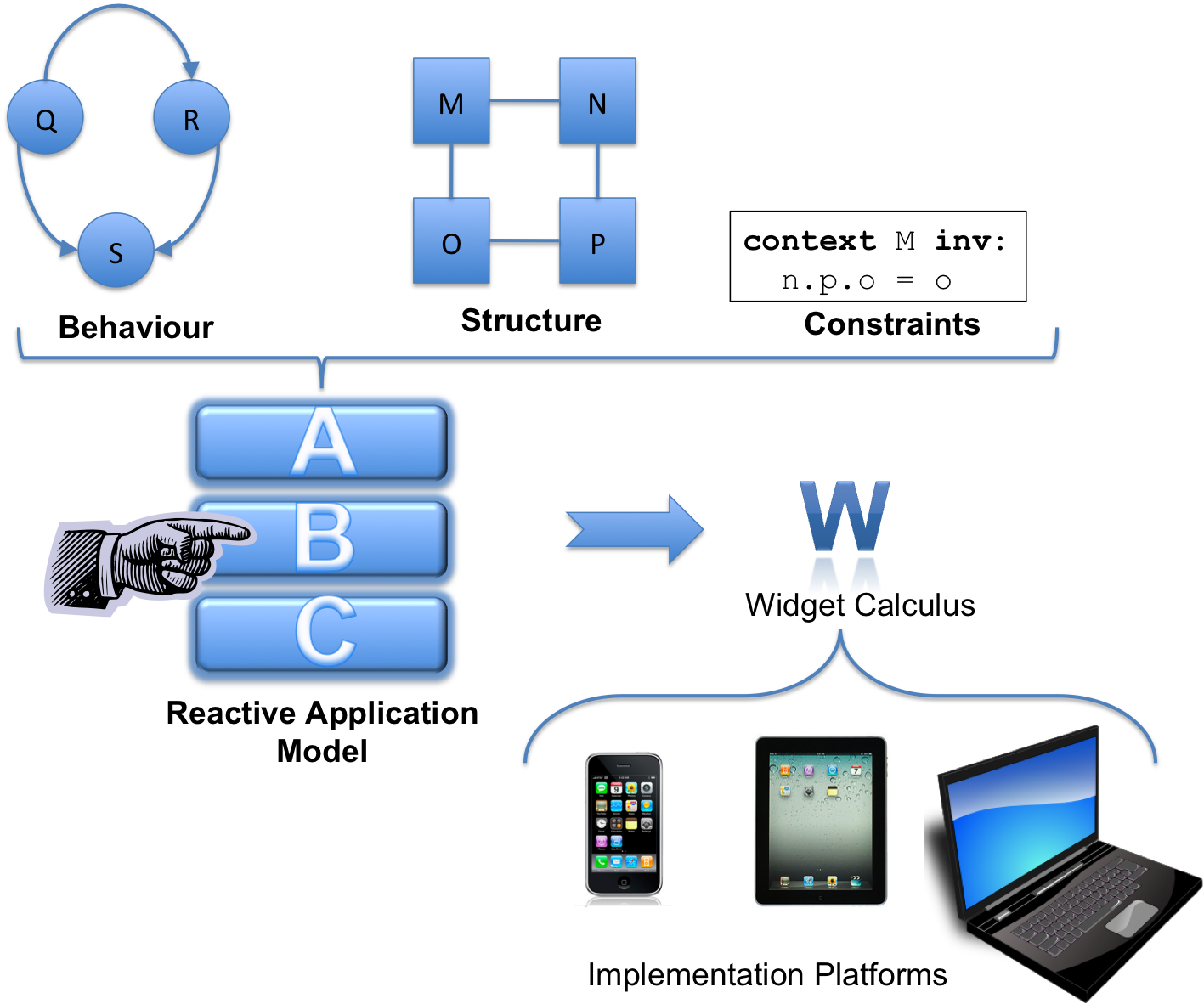}\hfill{}
\caption{Model Driven Reactive Applications\label{fig:CARA-Refinement}}
\end{figure}

The overall approach is shown in figure \ref{fig:CARA-Refinement}
where a reactive application model consists of structure, behaviour and some constraints.
A semantic mapping is used to translate the model to Widget where
it can be extended with detailed behaviour. An implementation mapping
is then viewed as a refinement of the semantics mapping in that it
translates the Widget program for an appropriate implementation platform.
The implementation mapping can be performed many different times for
the same Widget program in order to target multiple technologies.
An interpreter for the calculus has been written in Java and used to
implement Buddy against an external widget library written in Swing;
figures \ref{fig:Tony's-Phone}, \ref{fig:Tony-Knows-Sally} and
\ref{fig:Move-in-Range} are screen shots of the application.

The rest of the article is organized as follows: Section \ref{CARA} describes a typical \rapp case study called Buddy and performs domain-analysis in order to identify key features. Section \ref{sec:CARA-Models} introduces
a modelling language based on UML class diagrams and state machine that can be used to
represent application structure and behaviour; a model is given for Buddy. Section \ref{sec:The-Widget-Calculus}
introduces the Widget calculus and section \ref{sec:Semantic-Mapping} shows how \rapp models are translated
to Widget. Finally, section \ref{related_work} describes
related approaches and compares them to \rapp models and Widget.

\section{Reactive Applications}

\label{CARA}

Reactive applications have several common features. The user interacts
via a collection of screens and initiates computation by performing
actions that raise events and the application performs state transitions
in response to receiving events. This section provides a simple example
of a \rapp in section \ref{sub:Example-Application} and performs
domain analysis in section \ref{sub:Domain-Analysis} that identifies
the key common features. The screenshots in this section are taken from the 
prototype Widget interpreter with a Java Swing external widget interface and
where user interaction events have caused state transitions.

\subsection{Example Application: Buddy\label{sub:Example-Application}}

\begin{figure}[t]
\hfill{}\includegraphics[scale=0.9]{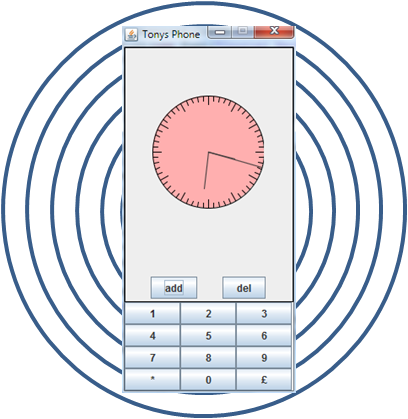}\hfill{}

\caption{Tony's Phone\label{fig:Tony's-Phone}}
\end{figure}

Figure \ref{fig:Tony's-Phone} shows a mobile phone. The phone is
always in contact with its network provider via a transmission cell
located nearby. Each phone has a unique address that is used by others
to contact the user, in this case it is \texttt{tony@widget.org}.

\begin{figure}
\hfill{}\includegraphics[scale=0.9]{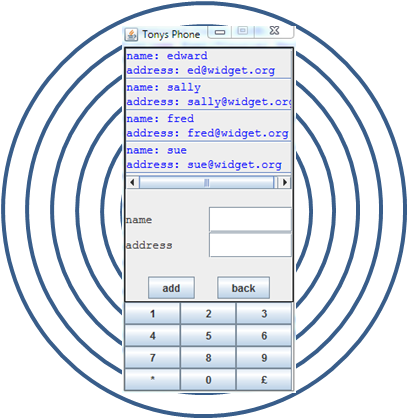}\hfill{}
\caption{Tony Knows Sally\label{fig:Tony-Knows-Sally}}
\end{figure}

Each phone contains a database of contacts. New contacts can be added
by clicking on the \texttt{add} button and entering the contact details.
As shown in figure \ref{fig:Tony-Knows-Sally}, Tony knows the address
of Sally. A new contact is added by clicking on the \texttt{add} button;
clicking on \texttt{back} returns to the previous screen.

\begin{figure*}[!]
\hfill{}\includegraphics[scale=0.8]{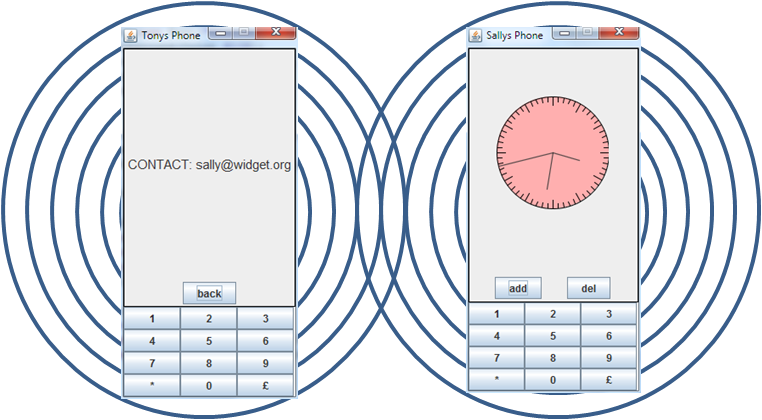}\hfill{}

\caption{Move in Range\label{fig:Move-in-Range}}
\end{figure*}

Multiple phones are always in contact with the service provider via
the local cell. Users want to know about contacts in their database
that are co-located. If Tony or Sally move within a predefined distance 
then Tony is informed as shown in figure \ref{fig:Move-in-Range}. 

To achieve this the service provider is told of the location of each phone; when one 
phone moves into the vicinity of the other then both phones are told of the 
availability of the other in terms of the contact address. If the address is in
the user's database then the phone flashes the contact.

This application has some key features. It is event driven where events arise 
either from the user (pressing a button)
or as changes in the context (buddy is in range). The application interfaces are simple
and organized hierarchically (for example the home screen contains a clock, a function button area and a numeric keypad). The application proceeds through a number
of states, driven by the events (the home state, the add-contact state, the buddy-alert state). The application has transient data (the values typed in the name and address fields in the add-contact screen) and persistent data (the list of contacts).

\subsection{Domain Analysis\label{sub:Domain-Analysis}}

A domain specific language is defined by performing a \emph{domain
analysis} \cite{Mernik:2005:DDL:1118890.1118892} on a target family
of applications in order to identify the common characteristic features.
The domain analysis leads to the design of a technology that conveniently
supports these features. Our domain analysis included working with
a media company to develop two iPhone applications: firstly to report Tour
de France cycle race results and secondly an on-line quiz. The domain 
analysis for \rapp identified the following key features:

\noindent \textbf{Screen Real Estate:} Different platforms make varying
amounts of screen available. For example a mobile platform is different
to a tablet which is different to a desktop browser. The standard
iPhone resolution is 480 by 320 pixel and the IPA supports a 1024
by 768 resolution. This compares to the Android screens, which vary
by hardware vendor but resolutions range to about 480 by 800 pixel.
However, in most cases the application logic is the same; how it is
realized in terms of the screen real estate can differ. Abstracting
away from the details of cross-platform differences is desirable when
maintaining a single application across multiple targets. 

\noindent \textbf{Layout Control:} Layout control is an important
consideration. Android controls layout through the use of XML files,
supporting different layout styles (linear, relative and absolute).
This compares to iPhone, that can do programmatic layout and XML type
interfaces using Interface Builder. Like screen size, it is desirable
to focus on application logic and factor out the layout control details
into external libraries.

\noindent \textbf{GUI Element Containership:} Most platforms use a
form of GUI element containership. In iPhone development, the emphasis
is on the application window with views and sub-views. 
These are then `stacked' onto each other to create structured interfaces.
Android uses a similar approach in terms of views and view-groups.
Interface control on both platforms have similarities and differences.
On the iPhone, views are normally controlled by the use of view controllers
that contain event handlers. In comparison, Android development uses
intents and activities. HTML structures interfaces in
terms of documents, tables, div's etc. This feature leads us to conclude
that a large number of \rapp GUIs can be expressed in terms of a tree
of \texttt{widgets} that manage data and behaviour and whose detailed
layout and rendering properties can be factored into platform specific
libraries. 

\noindent \textbf{Event Driven Applications}: Most mobile application
implementation languages register event handlers dynamically. Web
applications process events by dynamically testing identifiers embedded
in URLs. This method means there is a lack of checking at compile
time to prevent an application crashing. Contextual events such as 
platform orientation, GPS, and battery levels must be handled by a 
mobile application in suitable ways. This places a desirable feature 
requirement on development whereby the presence or otherwise of 
event handlers can be detected at compile-time.

\noindent \textbf{Hardware Features:} Modern day mobile devices come
equipped with many different features. These features include microphones,
accelerometers, GPS, camera, and close range sensors. These features
tend to be fairly standard in their behaviour if they are supported
by the platform. Although many platforms have comparable hardware
features, they differ in the details of how to control and respond
to them. \rapp development should allow the details of hardware to
be factored out into platform specific libraries whilst supporting
the events and controls associated with them.

\noindent \textbf{Object-Orientation:} Mobile and web-based applications
are typically OO. iPhone uses Objective-C and Android
uses Java. Javascript which is used by many web applications has an
object-oriented collection of data types for building applications.
Applications are built by constructing new and extending existing
class/object types.

\noindent \textbf{Transitional Behaviour: } \rapps\ 
execute in response to events that originate either from the user
or as context events from the platform. The application performs a
state transition in response to an event causing a change to the application's
state (or to a system that is connected to the application) and possibly
a new interface screen.

\noindent \textbf{Data Persistence: } \rapps\ usually need
to persist data to physical storage between application invocation.
Modern smartphone platforms currently have implementations of a SQLite,
a lightweight serverless single file database engine. 

\noindent \textbf{Contextual Events: }Within a mobile application,
not all events are directly invoked by the user. Mobile platforms
have to deal with event invocation from a range of different sources
based on its current contextual environment. For example, when the
battery is low on a phone normally the phone will display a message
to the user to recharge the battery. 

\noindent \textbf{Static Typing: }Type systems are used in programming
languages as a method of controlling legal and illegal program behaviour.
Static typing requires all type checking to be carried out during
run time, as opposed to dynamic typing that requires checking at run-time.
Since \rapps\ rely heavily on events and event handlers,
it is desirable that a program can be statically checked in order
to match handler definitions against all possible events that can
be raised.

\section{Reactive Models\label{sec:CARA-Models}}

Reactive models must support the key features that were identified in
\ref{sub:Domain-Analysis}. We use a DSL based on stereotyped UML
class diagrams to represent the structure of \rapp models and UML state
machine models to represent their behaviour. A {\it stereotype} is a tag of the form {\tt <<name>>} that is added to a standard UML element in order to designate it for a specific purpose. The tags are available to tools that process UML models so that they can take special action when generating code for example.

The stereotypes are used
by the semantic mapping to encode the structure into Widget and the
state machine is used to define Widget event handlers that make transitions
between screens. Section \ref{sub:Model-Features} describes the DSL
used for modelling, section \ref{sub:Phone-Structure} describes the
structure of the Buddy phone application and section \ref{sub:Phone-Behaviour}
describes its behaviour.

\subsection{Modelling Features\label{sub:Model-Features}}

A \rapp model is represented using a DSL for structure and a UML state
machine for behaviour. The structure of a \rapp is constructed
using widgets that generate and handle events. \emph{External} widgets,
represented as classes with stereotype \texttt{<\textcompwordmark{}<external>\textcompwordmark{}>},
are provided by the implementation platform. All widgets inherit from
the abstract external \texttt{Widget} class. \emph{User-defined} widgets
typically extend external widgets and are identified by the stereotype
\texttt{<\textcompwordmark{}<widget>\textcompwordmark{}>}.

Widgets define properties that are set when the widget is instantiated.
These are defined on a model using standard UML-style attributes.
Widgets may also define queries, events, commands and handlers. A
\emph{query} is an operation that can access the current state, but
cannot make any change to it; it is defined as a standard UML operation
on a class. An \emph{event} is a named, structured value that can
be raised by a widget. An external widget generates events as a result
of a change to the world state; user-defined widgets generate events
when they fail to handle an event generated by a widget they contain.
In addition user-defined widgets can explicitly generate events. Events
are defined as an operation with stereotype \texttt{<\textcompwordmark{}<event>\textcompwordmark{}>}.
A command is an operation that can change the program state; it is
specified using the \texttt{<\textcompwordmark{}<command>\textcompwordmark{}>}
stereotype; associations and properties can also be tagged as commands
when their values depend on the program state. A \emph{handler} is
a widget operation, tagged \texttt{<\textcompwordmark{}<handler>\textcompwordmark{}>},
used to handle events when they are raised. 

Widget references are defined using associations. Widget containment
hierarchies are modelled using UML black-diamond. Widget containment
is used to construct hierarchical GUIs and also to define how events
are handled. Events raised by a widget must be matched with a handler
with the corresponding signature (name and arguments). If the widget
raising the event defines such a handler then the event is supplied
to the handler that must produce a replacement for the widget in the
containment hierarchy. Otherwise, the widget does not define a handler
so the search continues with the parent. If the parent handles the
event then the parent is replaced in the containment hierarchy. This
process continues until a handler is found; static type checking guarantees
that a handler will be found for all events that can be raised.

The structure of a \rapp is a collection of state
models each of which must have a \emph{root container} widget. The
root is the GUI element that contains and references all other widgets
in that system state. A general purpose root container is the external
\texttt{Window} widget that contains GUI elements displayed on a phone
screen or a browser window. \texttt{Window} can be specialized to
produce application-specific external widgets.

\subsection{Phone Structure\label{sub:Phone-Structure}}

\begin{figure*}[t]
\hfill{}\includegraphics[scale=0.45]{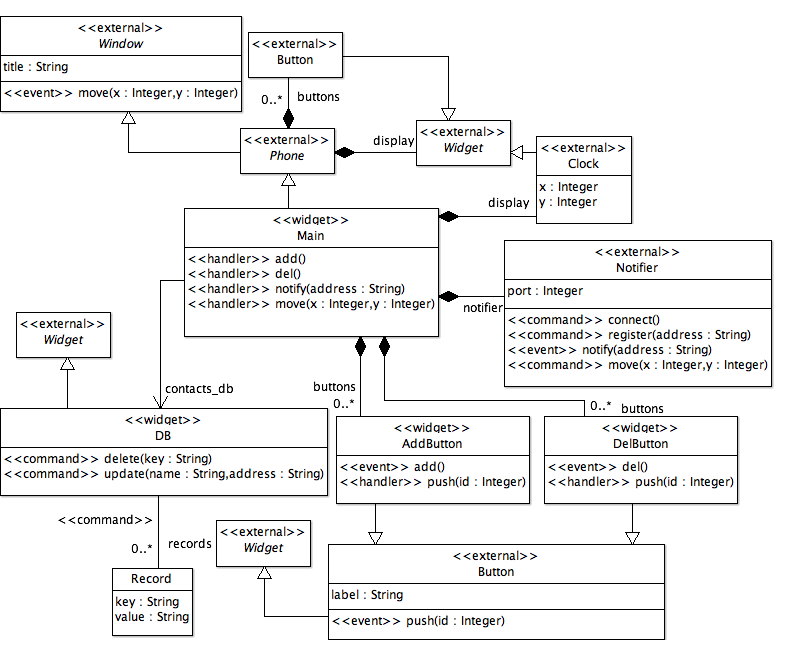}\hfill{}

\caption{Main Screen\label{fig:Main-Screen}}

\end{figure*}

The structure of the Buddy application is defined using four state
models that correspond to different screens. This section describes
three of these state models, the fourth is a simple variation and
so is omitted.

Figure \ref{fig:Main-Screen} shows the state model for the main screen.
All of the state models define root container widgets that extend
\texttt{Phone} which itself extends \texttt{Window}. The \texttt{Phone}
widget is external and must display a title, a display widget and
some buttons in an appropriate way that can differ between target
implementation platforms. In addition to the events generated by the
contained display and button widgets, \texttt{Phone} will generate
a \texttt{move} event since it inherits from \texttt{Window}.

The \texttt{Main} root container specializes the display and buttons
associations so that the main screen of the application presents a
clock and offers buttons for adding and deleting contacts. The \texttt{Clock}
widget is external and raises no events. Each button widget specializes
external \texttt{Button }that raises a \texttt{push} event (including
a unique numerical id) when pressed. Both \texttt{AddButton} and \texttt{DelButton}
have handlers for the \texttt{push} event that translate it into an
\texttt{add} and \texttt{del} event respectively. The \texttt{Main}
widget defines handlers for \texttt{add} and \texttt{del} that make
a transition to new application states as described below.

The \texttt{Main} widget also contains a \texttt{Notifier} that is
an external widget used to manage connections to the service provider.
Two commands, \texttt{connect} and \texttt{register}, are used to
initiate the connection with the provider after which \texttt{notify}
events will be generated when any phone that is connected to the same
provider comes into range. \texttt{Main} handles \texttt{move} events
that are passed on to the notifier whenever a phone moves from one
cell to another.

Each state in the model has a reference to a database widget \texttt{DB}.
The database widget manages a collection of records and provides commands
for deleting and modifying records in the database. Note that the
relationship between \texttt{Main} and \texttt{DB} is not containment
because the database does not generate any events.

\begin{figure*}[t]
\hfill{}\includegraphics[scale=0.45]{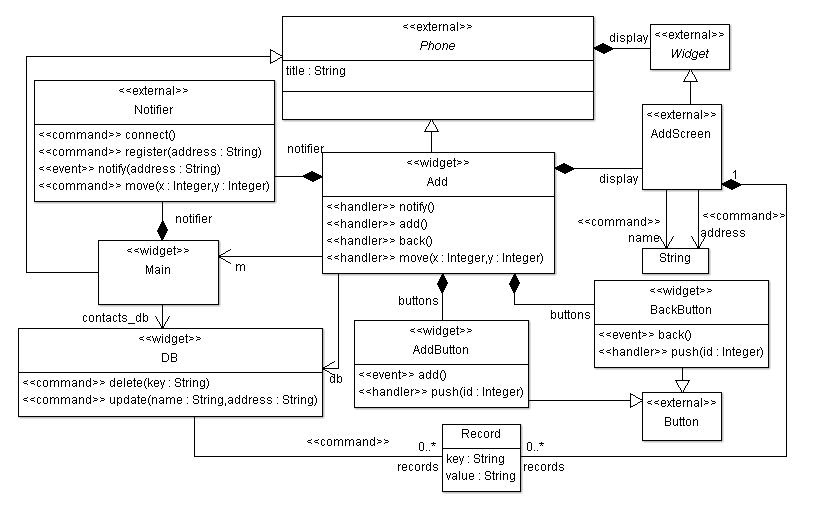}\hfill{}
\caption{Add New Contacts\label{fig:Add-New-Contacts}}
\end{figure*}

The system state used to add new contacts to the database is shown
in figure \ref{fig:Add-New-Contacts}. The root container is \texttt{Add},
the display is an external widget \texttt{AddScreen} that manages
browsing and adding new contact records. \texttt{AddScreen} provides
two commands that are used to yield name and address strings that
have been entered by the user via platform-specific text editing implemented
by the \texttt{AddScreen} widget. Buttons provided in the \texttt{Add}
state produce \texttt{add} and \texttt{back} events. The \texttt{add}
event updates the database before returning to the \texttt{Main} widget
and the \texttt{back} event just returns to the \texttt{Main} widget
as described in section \ref{sub:Phone-Behaviour}. 

Invariant constraints are expressed using OCL. The {\tt Add} widget
must share the database, title and notifier with the {\tt Main} widget:
\begin{lstlisting}
context Add inv:
  m.contacts_db = db and 
  m.notifier = notifier and
  m.title = title
\end{lstlisting}
The widget that
implements contact deletion is similar to \texttt{Add} and is therefore
not defined in this article.

\begin{figure}[t]
\hfill{}\includegraphics[scale=0.5]{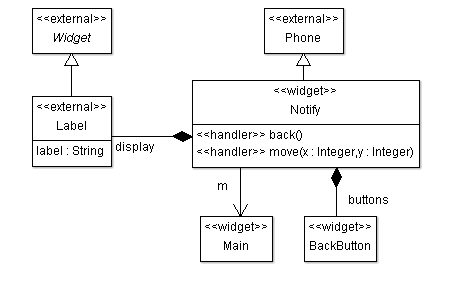}\hfill{}
\caption{Notify Screen\label{fig:Notify-Screen}}
\end{figure}

Figure \ref{fig:Notify-Screen} shows the widget for the notification
screen that occurs when a contact is detected in range. The display
is simply a label informing the phone user that the contact is nearby
and the button dismisses the notification and returns to the \texttt{Main}
screen. The title of the screens are the same:
\begin{lstlisting}
context Notify inv:
  m.title = title
\end{lstlisting}

\subsection{Phone Behaviour\label{sub:Phone-Behaviour}}

\begin{figure*}
\hfill{}\includegraphics[scale=0.5]{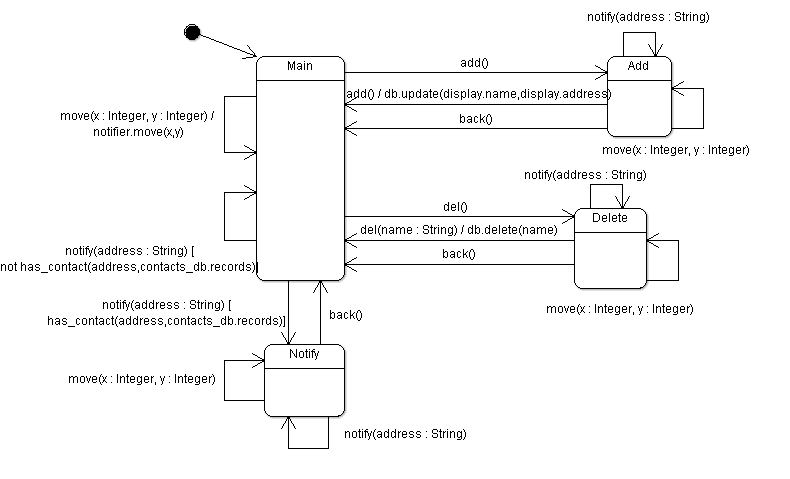}\hfill{}
\caption{State Transitions\label{fig:State-Transitions}}
\end{figure*}

The behaviour model for the phone application is shown in figure
\ref{fig:State-Transitions}. Each state corresponds to a root container
widget. Transitions correspond to the events that are handled by a
root container. The figure shows that the application starts in the
\texttt{Main} state. Pressing the \texttt{add} or \texttt{del} buttons
cause a corresponding state transition. Notification events are ignored
unless they occur in the \texttt{Main} state; if the contact is known
then they are displayed via the \texttt{Notify} state, otherwise the
event is ignored.

\section{The Widget Calculus\label{sec:The-Widget-Calculus}}

\begin{figure}
\begin{center}
\begin{minipage}{\columnwidth}
\begin{lstlisting}[numbers=right,basicstyle={\ttfamily\scriptsize},deletekeywords={type}]
e ::=                                 expressions
  |   x                               variables
  |   k                               constants
  |   [e,...]                         lists
  |   {x=e;...}                       records
  |   e.x                             field refs   
  |   fun(x:t,...):t e                functions
  |   e(e,...)                        applications
  |   if e then e else e              conditionals
  |   fix(e)                          fixed points
  |   raise x(e,...)                  events
  |   do {x:t<-e;...; return e}       blocks
  |   widget x:t (e) {x:t<-e;...}     widgets
  |   top                             universal
  |   Fun[x,...] e                    type abstraction
  |   e[t,...]                        instantiate type
z ::= x(t,...)                        events
v ::=                                 values
  |   x                               variables
  |   k                               constants
  |   [v,...]                         lists
  |   {x=v;...}                       records
  |   fun(x:t,...):t e                functions
  |   raise x(v,...)                  exceptions
  |   do {x:t<-e;...; return e}       blocks
  |   widget x:t (e) {x:t<-e;...}     widgets
  |   top                             universal value
\end{lstlisting}
\end{minipage}
\end{center}
\caption{Widget Expressions, Commands, Values\label{fig:Widget-Expressions}}

\end{figure}

The Widget Calculus is a simple functional language that has been
designed to support \rapp programs. Widget is both stand-alone and
can be used as the target of a \rapp PIM. In addition to being a standard functional
language, Widget provides three key \rapp elements: \emph{widgets}
that encode sources of reactive behaviour including both externally
defined widgets and user-defined widgets; \emph{commands} that make
changes to the current world-state; \emph{events} that arise from
state-changes including both externally generated events and user-defined
events. The core syntax of Widget-expressions is defined in figure
\ref{fig:Widget-Expressions}. The rest of this section describes
features of the syntax and concludes with an informal description
of its operational semantics.

\subsection{Basic Features}

Widget consists of standard functional language expressions defined
in figure \ref{fig:Widget-Expressions}: variables (2), strings (3),
numbers (3), booleans (3), lists (4), records (5), field references
(6). Functions (7) include the types of the arguments and the return
type. Applications (8) and conditionals (9) are standard. A recursive definition
is created using a fixed-point expression (10) in the usual way such that
\texttt{f(fix(f))=fix(f)} (syntactic sugar is used to define mutually recursive
local definitions as {\tt\bf letrec} using {\tt fix} in the usual way). 
Mutually recursive top-level definitions are 
introduced by keywords {\tt fun}, {\tt val}, {\tt type}.

\subsection{Polymorphism}

Widget is a statically-typed language and supports operators that
construct external widgets (see below). An example of such an operator
is {\tt db} that creates a database widget that maps keys to values, and
that implements commands to update and access the database. The behaviour 
of a database is independent of the actual types of the keys and values,
therefore the constructor {\tt db} is {\it polymorphic}. In order for 
Widget to statically check the use of databases, the key and values types
must be supplied to the constructor: {\tt db[int,str]} or {\tt db[str,int]}
etc.

Programs that work uniformly over all types (such as many list processing
functions) can be declared with respect to one or more type parameters
(15). When an expression that is parametric in one or more types is
used, the actual type is supplied as an argument (16). A standard
example is the identity function declared as: \texttt{id=Fun{[}t{]}
fun(x:t):t x} and then used in two different ways: \texttt{id{[}int{]}(10)}
and \texttt{id{[}bool{]}(true}). A special constant is used for the
empty list: \texttt{{[}{]}{[}t{]}} in order to declare the type of
the elements; therefore the empty list of integers \texttt{{[}{]}{[}int{]}
}is different to the empty list of strings \texttt{{[}{]}{[}str{]}}.

\subsection{Commands}

Commands are values that can be used to query the program state or
to change program state or both. The program state includes the current collection of widgets, therefore there are commands that create new widgets (we assume inaccessible widgets are garbage collected). The details of the program state depends on the set of imported external widgets that are used, for example an external widget that implements a database will have state that is modified by adding and removing elements. 

The underlying platform may also generate events that must be handled by user code. For example an external widget that manages the battery in a mobile phone may generate events when the amount of charge reduces below a preset level. User defined commands may choose to handle an event or to promote the event to a surrounding handler. Therefore commands can also generate events.

Widget commands are values that can be passed as values and returned as results. Therefore a command expression evaluates to produce a command. A command expression has the type {\tt<t>} and we say that the denoted command is \emph{performed} to \emph{yield} a value of type {\tt t}. Typically command expressions are used as follows: to express a user-defined widget; to initialize the fields of a user-defined widget; to define the body of a handler in a user-defined widget. The latter is interesting because user-defined widgets have fields that can hold any type of value, therefore a field that contains a function whose body is a command-expression is equivalent to a method in an object-oriented programming language.

Widget has some built-in commands and the do-expression (12) that builds composite commands. Locations containing values of type {\tt t} have a type {\tt !t}. The following builtin operators deal with locations: {\tt loc} has type {\tt Forall(t)(t)-><!t>}, {\tt get} has type {\tt Forall(t)(!t) -><t>} and {\tt set} has the type {\tt Forall(t)(!t,t)-><t>}. The following function maps locations to commands that add 1 to the contents of the location and yields the new value:

\begin{lstlisting}
fun add1(l:!int):<int> = do { 
  x:int <- get[int](l); 
  y:int <- set[int](l,x+1) 
  return y 
}
\end{lstlisting}

\noindent Commands are first-class values that can be passed as arguments and
returned as results. Commands can also be nested as shown in the following 
example:

\begin{lstlisting}
fun add2(l:!int):<int> = do { 
  void:int <- add1(l); 
  z:int <- add1(l); 
  return z 
}
\end{lstlisting}

\noindent Events are generated by a \texttt{raise} command (11) that,
when performed, yields a distinguished value \texttt{{*}}, and is
processed as described below.

\subsection{Widgets}

\label{widgets}

A widget definition (13) is a command that yields a new widget. Each
widget has the following form:

\begin{lstlisting}
widget self:t (parent) { 
  x1:t1 <- e1; 
  ...; 
  xm:tm <- em 
}
\end{lstlisting}
\noindent
where \texttt{self} is the name that can be used in the body of the
widget to refer to itself and may be omitted if not used. All widgets
inherit from a \texttt{parent} widget supplied as a command.
The special command \texttt{top} (14) is used to create
a distinguished widget that has no parent and acts like \texttt{Object}
in object-oriented programming languages. 

The idea is that user-defined widgets ultimately inherit from external
widgets. The external widget will generate an event that the child
can handle via its components. If the user-defined widget does not define
any components then it is equivalent to the parent:
\begin{lstlisting}[mathescape=true]
widget (p) {} $\equiv$ p 
\end{lstlisting} 
\noindent
Each definition \texttt{x:t <- e} in the body of the widget defines
a command \texttt{e} that yields a component. The component is named
\texttt{x} and can be referenced within other definitions and the parent. 
We use the convention that definitions whose value is a function can define
the function in-line and that \texttt{x:t = e} is equivalent to \texttt{x:t
<- do \{ return e \}}.

A widget may define any type of component, but typically contains
widgets and functions. The contained widgets raise events some of
which may be handled by the container's functions. The scope of variables
in widget body definitions are scoped so that names used earlier in the list are scoped over values
later in the list except for function definitions that are only available as
event handlers. Therefore, value and function definitions in widgets can be treated separately 
by re-ordering values before functions in the body. Furthermore, it is possible to simplify
any definition using the following equivalence:
\begin{lstlisting}[mathescape=true]
widget(e) { x <- e; d } $\equiv$ 
widget(widget(e) { x <- e }) { d }
\end{lstlisting}
Each handler function
must return a command that yields a widget. For example, if a window
contains a single button that does nothing when it is pressed then
we construct a widget with a parent using the external
constructors {\tt window} and {\tt button}:
\begin{lstlisting}
widget 
    self:MyWindow 
    (window('My Window',button('PUSHME'))) {
  push(id:int):<MyWindow> = do { return self }
}
\end{lstlisting}
In the widget above, the parent is a window with a title {\tt 'My Window'}.
The contents of the window is the button {\tt button('PUSHME')}. When a button
is selected, it generates events of the type {\tt push(int)}. Since the
widget defines a handler whose signature matches the event then the handler
defines a replacement for the entire window when the button is pressed. The
handler returns a command that yields {\tt self}, causing the window to be
replaced with itself, i.e. nothing happens when the button is pressed.

Equivalently, the handler can be processed by the component widget. In the following, the button {\tt b} handles the {\tt push} event; the button is replaced with itself {\it inside} the window:
\begin{lstlisting}
widget(window(title,b)) {
  title:str = 'My Window';
  b:MyButton <- 
    widget self:MyButton (button('PUSHME')) {
      push(id:int):<MyButton> = do { 
        return self 
      }
    }
}
\end{lstlisting}
The following is a window that oscillates between two buttons when they are pressed:
\begin{lstlisting}
widget(window('MyWindow',b1)) {
  b1:MyButton <- 
    widget(button('FORWARD')) { 
      push(i:int):<MyButton> = do { 
        return b2 
      } 
    };
  b2:MyButton <- 
    widget(button('BACK')) { 
      push(i:int):<MyButton> = do { 
        return b1 
      } 
    }
}
\end{lstlisting}

\subsection{Two Examples}

A Widget program cycles through the following phases: 
\begin{description}
\item[\it reducing] an expression to produce a command. This involves applying operators to operands and the construction of basic data structures. Reduction is side-effect free.  
\item[\it performing] a command with respect to the state of a root widget. The command can change the state of the widget tree and its context. For example, a command allocates unique identifiers to widgets or calls a system library to update or access a local database. 
\item[\it displaying] a widget tree in a technology-specific way and waiting for an event. The event may originate from a user interaction with the system or may originate from the system context. In all cases an event can be associated with a unique widget in the tree.
\item[\it processing] an event by delivering it to a uniquely identified widget (the {\it receiver}). The event names a handler in the receiver whose body is produces a new expression ready for a fresh reduction step. The reduction produces a command that is performed to produce a replacement widget for the receiver.
\end{description}
This section contains two simple examples that show how these phases operate in terms of the calculus. 

\noindent{\bf Example 1:} A screen widget contains a single button that displays a label {\tt PUSHME}. Pushing the button is an identity step on the system. The starting expression is:
\begin{lstlisting}[basicstyle={\ttfamily\scriptsize},morekeywords={widget,do,return},keywordstyle=\bf]
letrec
  main = 
    widget self (screen(50,50,50,50,push)) { 
      move(x,y) = do { return self } 
    };
  push = 
    widget self (button('PUSHME')) {
      push(i) = do { return self }
    }
in main
\end{lstlisting}
After reduction we get the following command:
\begin{lstlisting}[basicstyle={\ttfamily\scriptsize},morekeywords={widget,do,return},keywordstyle=\bf]
widget self (screen(50,50,50,50,
  widget self (button('PUSHME')) {
    push(i) = do { return self }
  }) { 
  move(x,y) = do { return self } 
}
\end{lstlisting}
The command is performed by allocating unique identifiers to each widget in the tree. The built-in widgets {\tt screen} and {\tt button} are allocated identifiers {\tt 2} and {\tt 0} respectively. The user defined widget identifiers are shown in parentheses after the keyword {\tt\bf widget}:
\begin{lstlisting}[basicstyle={\ttfamily\scriptsize},morekeywords={widget,do,return},keywordstyle=\bf]
widget(3) self (screen(2,50,50,50,50,
  widget(1) self (button(0,'PUSHME')) {
    push(i) = do { return self }
  }) { 
  move(x,y) = do { return self } 
}
\end{lstlisting}
The {\tt displaying} phase then displays the tree as a screen containing a button. An event is receieved when the user presses the button with identifier {\tt 0}. This is denoted as {\tt <w>i<-push(i)} where the argument to {\tt push} is the source identifier (in case handlers are shared between different widgets). 

{\it Processing} an event traverses the widget {\tt w} until the identifier {\tt i} is found. When the widget with identifier {\tt i} is found we will retain the context so that it can be replaced:
\begin{lstlisting}[basicstyle={\ttfamily\scriptsize},morekeywords={widget,do,return},keywordstyle=\bf]
<widget(3) self (screen(2,50,50,50,50,
  widget(1) self (button(0,'PUSHME')) {
    push(i) = do { return self }
  } { 
  move(x,y) = do { return self } 
}>0<-push(0)
\end{lstlisting}
Widgets {\tt 3} and {\tt 2} are incorrect so the search moves down the tree:
\begin{lstlisting}[basicstyle={\ttfamily\scriptsize},morekeywords={widget,do,return},keywordstyle=\bf]
widget(3) self (screen(2,50,50,50,50,
  <widget(1) self (button(0,'PUSHME')) {
    push(i) = do { return self }
  }>0<-push(0))) {
  move(x,y) = do { return self } 
}
\end{lstlisting}
The widget with identifier {\tt 1} {\it contains} the widget with identifier {\tt 0}. A {\tt button} widget is external and does not define any handlers, therefore its inner-most container that defines a handler with the appropriate name will handle the event. The body of the handler is an expression that is reduced to produce a command (denoted using {\tt <} and {\tt >}):
\begin{lstlisting}[basicstyle={\ttfamily\scriptsize},morekeywords={widget,do,return},keywordstyle=\bf]
widget(3) self (screen(2,50,50,50,50,
  <do { 
    return widget(1) self (button(0,'PUSHME')) {
      push(i) = do { return self }
    }
  }>) { 
  move(x,y) = do { return self } 
}
\end{lstlisting}
Performing the command returns a widget that is used as a replacement for the receiver. Since the command returns the receiver (via {\tt self}) this is an identity step:
\begin{lstlisting}[basicstyle={\ttfamily\scriptsize},morekeywords={widget,do,return},keywordstyle=\bf]
widget(3) self (screen(2,50,50,50,50,
  widget(1) self (button(0,'PUSHME')) {
    push(i) = do { return self }
  }) { 
  move(x,y) = do { return self } 
}
\end{lstlisting}
The tree is now ready to receive further events and the process loops indefinitely.

\noindent{\bf Example 2:} Our second example shows how multiple events are handled. Of course since events originate from user interaciton, they are serialized, however they may target different GUI widgets. The following example shows how a button toggles between two states. The following mutually recursive definitions create a screen:
\begin{lstlisting}[basicstyle={\ttfamily\scriptsize},morekeywords={widget,do,return},keywordstyle=\bf]
letrec
  main = 
    widget self (screen(50,50,50,50,push)) { 
      move(x,y) = do { return self } 
    };
  push = 
    widget (button('PUSHME')) {
      push(i) = do { return pushed }
    };
  pushed = 
    widget (button('PUSHED')) {
      push(i) = do { return push }
    }
in screen
\end{lstlisting}
They evauate to produce a command:
\begin{lstlisting}[basicstyle={\ttfamily\scriptsize},morekeywords={widget,do,return},keywordstyle=\bf]
widget self (screen(50,50,50,50,push)) { 
  move(x,y) = do { return self } 
}
\end{lstlisting}
The command is performed to allocate unique identifiers:
\begin{lstlisting}[basicstyle={\ttfamily\scriptsize},morekeywords={widget,do,return},keywordstyle=\bf]
widget(5) self (screen(4,50,50,50,50,
  widget(3) (button(0,'PUSHME')) {
    push(i) = do { 
      return widget(2) (button(1,'PUSHED')) {
        push(i) = do { return push }
      } 
    }
  })) { 
  move(x,y) = do { return self } 
}
\end{lstlisting}
The event {\tt 0<-push(0)} is received and targets the appropriate widget:
\begin{lstlisting}[basicstyle={\ttfamily\scriptsize},morekeywords={widget,do,return},keywordstyle=\bf]
widget(5) self (screen(4,50,50,50,50,
  <widget(3) (button(0,'PUSHME')) {
    push(i) = do { 
      return widget(2) (button(1,'PUSHED')) {
        push(i) = do { return push }
      } 
    }
  }.push(0)>)) { 
  move(x,y) = do { return self } 
}
\end{lstlisting}
It is handled and produces a command that, when performed, produces a replacement widget for {\tt 3}:
\begin{lstlisting}[basicstyle={\ttfamily\scriptsize},morekeywords={widget,do,return},keywordstyle=\bf]
widget(5) self (screen(4,50,50,50,50,
  <do { 
    return widget(2) (button(1,'PUSHED')) {
      push(i) = do { return push }
  }>)) { 
  move(x,y) = do { return self } 
}
\end{lstlisting}
The result is a new screen where the button has changed state:

\begin{lstlisting}[basicstyle={\ttfamily\scriptsize},morekeywords={widget,do,return},keywordstyle=\bf]
widget(5) self (screen(4,50,50,50,50,
  widget(2) (button(1,'PUSHED')) {
    push(i) = do { return push }
  })) { 
  move(x,y) = do { return self } 
}
\end{lstlisting}
The event {\tt 1<-push(1)} causes an equivalent sequence of changes to occur, resulting in the original system state:
\begin{lstlisting}[basicstyle={\ttfamily\scriptsize},morekeywords={widget,do,return},keywordstyle=\bf]
widget(5) self (screen(4,50,50,50,50,
  widget(3) (button(0,'PUSHME')) {
    push(i) = do { 
      return widget(2) (button(1,'PUSHED')) {
        push(i) = do { return push }
      } 
    }
  })) { 
  move(x,y) = do { return self } 
}
\end{lstlisting}

\begin{figure}
\begin{center}
\begin{minipage}{\columnwidth}
\begin{lstlisting}[basicstyle={\ttfamily\scriptsize},numbers=right,deletekeywords={top,type}]
t ::= str                                 strings
  |   int                                 integers
  |   bool                                booleans
  |   [t]                                 lists
  |   {x:t;...}                           records
  |   <t> raises z,...                    commands
  |   t + t                               union
  |   Widget(t) raises z,... {x:t;...}    widgets
  |   Top                                 top
  |   (t,...)->t                          functions
  |   x[t,...]                            application
  |   rec x.t                             fixed points
  |   x                                   type variable
  |   {{x=t;...}}                         type record
  |   t.x                                 type ref
  |   *                                   unit
  |   Forall(x,...)t                      universal
  |   !t                                  locations
\end{lstlisting}
\end{minipage}
\end{center}
\caption{Widget Types\label{fig:Widget-Types}}
\end{figure}

\begin{figure*}
\begin{center}
\begin{tabular}{lr}
\inference[T-VAR]{}{\Gamma[x\mapsto\alpha]\vdash x:\alpha}
&
\inference[T-TRUE]{}{\Gamma\vdash\mathrm{\tt true}:\mathrm{\tt bool}}
\\\\
\inference[T-FUN]{\Gamma{[x_i\mapsto\alpha_i]}^{i\in[0,n]}\vdash t:\alpha}{\Gamma\vdash\mathrm{\tt fun}(x_i^{i\in[0,n]})t:(\alpha_i^{i\in[0,n]})\rightarrow\alpha}
&
\inference[T-APP]{\Gamma\vdash t:(\alpha_i^{i\in[0,n]})\rightarrow\alpha\\\Gamma\vdash t_i:\alpha_i\ \forall\ i\in[0,n]}{
\Gamma\vdash t(t_i^{i\in[0,n]}):\alpha}
\\\\
\inference[T-IF]{\Gamma\vdash t_1:\mathrm{\tt bool}\\\Gamma\vdash t_2:\alpha\\\Gamma\vdash t_3:\beta}{\Gamma\vdash\mathrm{\tt if}\ t_1\ \mathrm{\tt then}\ t_2\ \mathrm{\tt else}\ t_3:\alpha\oplus\beta}
&
\inference[T-REC]{\Gamma\vdash t_i:\alpha_i\ \forall\ i\in[0,n]}{\Gamma\vdash\{{x_i=t_i}^{i\in[0,n]}\}:\{{x_i:\alpha_i}^{i\in[0,n]}\}}
\\\\
\inference[T-REF]{\Gamma\vdash t:\{{x_i:\alpha_i}^{i\in[0,n]}\}}{\Gamma\vdash t.x_k:\alpha_k\ \exists k\in[0,n]}
&
\inference[T-LIST]{\Gamma\vdash t_i^{i\in[0,n]}:\alpha}{\Gamma\vdash[t_i^{i\in[0,n]}]:[\alpha]}
\\\\
\inference[T-OPT-1]{\Gamma\vdash t:\alpha}{\Gamma\vdash t:\alpha+\beta}
&
\inference[T-OPT-2]{\Gamma\vdash t:\beta}{\Gamma\vdash t:\alpha+\beta}
\\\\
\inference[T-FIX]{\Gamma\vdash e:(t)\rightarrow t}{\Gamma\vdash\mathrm{\tt fix}(e):t}
\\\\
\multicolumn{2}{c}{
\inference[T-EXC]{\Gamma\vdash e_i:t_i\ \forall i\in[0,n]}{\Gamma\vdash\mathrm{\tt raise}\ {x(e_i)}^{i\in[0,n]}:\left\langle {*}\right\rangle \uparrow\{{x(t_i)}^{i\in[0,n]}\}}}
\\\\
\multicolumn{2}{c}{
\inference[T-DO]{
\Gamma[{x_j\mapsto t_j}^{j\in[0,i-1]}]\vdash e_i:<t_i>\uparrow X_i\ \forall i\in[0,n]\\
\Gamma[{x_i\mapsto t_i}^{i\in[0,n]}]\vdash e:t}
{\Gamma\vdash\mathrm{\tt do}\ \{ {x_i:t_i\leftarrow e_i}^{i\in[0,n]}\ \mathrm{\tt return}\ e\}:\left\langle t\right\rangle \uparrow{\displaystyle\bigcup_{i\in[0,n]}} X_i}}
\\\\ 
\inference[T-TOP]{}{\Gamma\vdash\mathrm{\tt top}:\mathrm{\tt Top}}
&
\inference[T-UNI-I]{\Gamma\vdash e:t{[x_i\mapsto t_i]}^{i\in[0,n]}\ \mathrm{any}\ t_i\not\in\mathrm{FV}(t)}{\Gamma\vdash \mathrm{\tt Fun}[{x_i^{i\in[0,n]}}]e:\mathrm{\tt Forall}[{x_i}^{i\in[0,n]}]t}
\\\\

\inference[T-UNI-E]{\Gamma\vdash e:\mathrm{\tt Forall}[{x_i^{i\in[0,n]}}]t}{\Gamma\vdash e[{t_i^{i\in[0,n]}}]:t[{x_i\mapsto t_i}^{i\in[0,n]}]}
\\\\
\multicolumn{2}{c}{
\inference[T-WID1]{%
\Gamma\vdash e:\left\langle W(X)\right\rangle\uparrow X'\\
\Gamma[x\mapsto t]\vdash e_i:\left\langle(\tilde{t_i})\rightarrow\left\langle t_i\right\rangle\uparrow X_i\right\rangle\uparrow X_i'\ \forall i\in[0,n]\\
t=\mathrm{\tt Widget}(W(X))\uparrow (X\cup X_i^{i\in[0,n]})-{\{x_i(\tilde{t_i})\}}^{i\in[0,n]} {\{x_i:(\tilde{t_i})\rightarrow\left\langle t_i\right\rangle\uparrow X_i\}}^{i\in[0,n]}
}{%
\Gamma\vdash\mathrm{\tt widget}\ x:t(e){\{x_i:(\tilde{t_i})\rightarrow\left\langle t_i\right\rangle\uparrow X_i=e_i\}}^{i\in[0,n]}:\left\langle t\right\rangle\uparrow X'\cup X_i'^{i\in[0,n]}
}}
\\\\
\multicolumn{2}{c}{
\inference[T-WID2]{%
\Gamma[x\mapsto W(X)]\vdash e':\left\langle W'(X')\right\rangle\uparrow X''\\
\Gamma[x'\mapsto t]\vdash e:\left\langle W(X)\right\rangle\uparrow X'''\\
t=\mathrm{\tt Widget}(W(X))\uparrow X\cup X' \{x:W(X)\}
}{%
\Gamma\vdash\mathrm{\tt widget}\ x:t(e') \{ x:W(X)\leftarrow e\}:\left\langle t\right\rangle\uparrow X''\cup X'''
}}
\end{tabular}
\end{center}
\caption{Type Relation}
\label{type_relation}
\end{figure*}

\subsection{Types}
\label{types} 

Reactive applications execute in terms of function application, message passing and by handling events. Some
implementation technologies such as Javascript, are dynamically typed, and others, such as Java for Android, 
are statically typed. In virtually all cases, events are handled by registering event handlers with the
underlying framework so that the handler is called when the event occurs. Handler registration occurs at 
run-time; in such languages it is not possible to statically analyze an application for the existence of all required handlers.

Widget is a strongly typed, statically checked language. In addition to statically checking that the types of
operator arguments and field references are correct, Widget can check that all possible events raised by an 
application have an appropriate handler definition. For context aware applications this means that a tool
can check that all situations are handled, for example low battery power, change of platform orientation, etc.
Whilst this does not guarantee the the application is correct, it reduces the possibility that the developer 
has inadvertently omitted a handler definition that could lead to sub-optimal or inappropriate application behaviour.

The types are defined in figure
\ref{fig:Widget-Types}. Constants are strings, integers or booleans
(1-3), a list must contain elements of a single type (4), the types
of each field in a record may be different (5), a command that yields
a value of type \texttt{t} is of type \texttt{<t>} (6), a value of
a union type (7) is a value of either component type, a widget expression
is a command that yields a value of a widget type (8) and each component
field in the widget expression must be a command that yields a value
of the corresponding field type, the expression \texttt{top} is a
command that yields a value of type \texttt{Top} (9), a function has
a function type (10), a universal type (17) can be applied to type
arguments to yield a type (11), a type may be recursive (12), types
may be bound to type variables (13), types may be packaged up into
type records (14) and referenced (15), finally, an event raising expression
is a command that yields the value of type \texttt{{*} }(16)\texttt{.}

Figure \ref{type_relation} defines a relation between type assignments $\Gamma$,  
expressions $e$ and types $t$ such that $\Gamma\vdash e:t$ holds when $e$ is
assigned type $t$ when free variables in $e$ are assigned types by $\Gamma$. The
relation is standard except in terms of event handling where $\mathrm{\tt raises} X$ is
written $\uparrow X$ for brevity. T-IF combines the types of the consequent and
alternative arms $\alpha\oplus\beta$ where $(\left\langle t\right\rangle \uparrow X)\oplus(\left\langle t'\right\rangle \uparrow X')=\left\langle t+t'\right\rangle \uparrow X\cup X'$, otherwise $\oplus$ is the same as $+$. T-EXC defines a
{\tt raise} command to yield the unit value and to raise an event of the appropriate
type. T-DO combines all events raised by the definitions. 

The refactoring of widget expressions in section \ref{widgets}, where single value 
definitions are extracted the parent, allows us to  define widget type assignment as two
separate rules T-WID1 and T-WID2. The shorthand $W(X)$ is used for 
{\tt Widget(t) raises X { d }} where only the events $X$ raised by the widget type are of interest.
T-WID1 defines type assignment where the body of a widget consists of handler definitions;
the events handled by the child are erased from those raised by the parent. In
T-WID2 the events raised by the contained widget are added to those raised by the parent.

\begin{figure}
\hfill{}\includegraphics[scale=0.4]{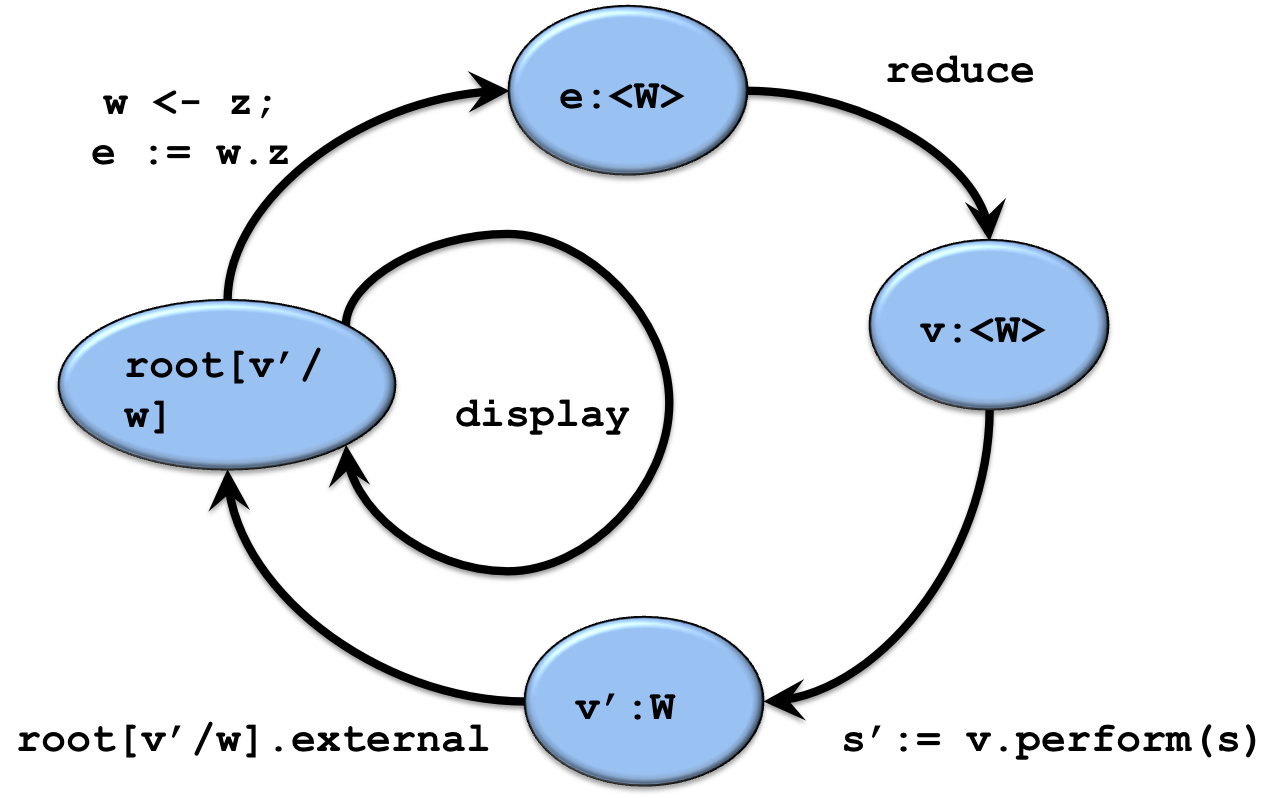}\hfill{}

\caption{Execution Cycle\label{fig:Execution-Cycle}}
\end{figure}

\subsection{Operational Semantics}

In order to run on a platform, a Widget program must have a specific type: $\left\langle W(\emptyset)\right\rangle\uparrow\emptyset$, i.e. a command that yields a widget whose events
have all been erased. Program execution cycles through four stages: evaluation; commands; display; event handling. 
Firstly the program is evaluated, or {\it reduced}, to produce a value, or {\it normal form}. Given the type restrictions, 
the value is a command that yields a widget as a result of the second stage of execution. The
second stage can perform side-effects. 

A widget is a tree $t$ whose leaves are external widgets
(or {\tt top}); the tree is projected onto a tree $t'$ of external widgets that is displayed 
on an implementation platform. The third stage of execution displays $t'$.
At this point the application waits to receive an event $e$, either from the underlying platform
(a context event) or from user interaction. Stage four involves handling $e$ by mapping from the
receiving external widget in $t'$ to the corresponding widget $w$ in $t$. By 
traversing from $w$ to the root of $t$ a {\it most specific handler} $h$ is found. The body of 
$h$ is an expression $e$ of type $\left\langle W'(X)\right\rangle\uparrow X'$ that yields a replacement
for $w$. Since $e$ is of the appropriate type, evaluation can re-start from stage one.

The operational semantics of Widget programs is shown as a state-machine
in figure \ref{fig:Execution-Cycle}. The root container widget is
called \texttt{root} and on the first iteration the starting state
shown at the top of the diagram is \texttt{e=root=w} and \texttt{root:<W>}
where \texttt{W} is a sub-type of \texttt{Window}. Reduction produces a command \texttt{v:<W>}. The
command is performed with respect to the program state \texttt{s }to
yield a value \texttt{v'} and a new state \texttt{s'}. The value \texttt{v'}
is the replacement for the current target of the event: \texttt{w
}in \texttt{root} (in the first case this replaces the root with itself).
The root containment tree is projected onto a tree of builtin widgets
using \texttt{external} which is then displayed on a screen. The system
then waits for an event\texttt{ z} which is sent to a widget \texttt{w}
that is contained in \texttt{root}. At this point the target widget
\texttt{e }is reset to the body of the handler for \texttt{z }in \texttt{w}.
The cycle continues, each time, the target of an event is replaced
by the value yielded by the body of the handler for the event.

\section{Mapping and Behaviour\label{sec:Semantic-Mapping}}

Section \ref{sec:CARA-Models} has described how \rapp models can represent
a mobile phone application. Section \ref{sec:The-Widget-Calculus}
has described a technology-independent language for representing reactive
applications. The \rapp modelling language could be translated directly
onto an implementation technology. In practice, this is how things
would be done; however, such a strategy leads to a semantic definition
for the application in terms of an implementation platform. This strategy
has two significant disadvantages: firstly implementation technologies
tend to be complex; secondly if the application is to be realized
across multiple platforms then it is much more attractive to use a
technology-independent semantic domain.

Our hypothesis is that Widget provides a suitable precise, lightweight
and executable platform for \rapps. This section describes a
translation from \rapp models to Widget programs in terms of the Buddy
case study.

\subsection{General Mapping}

The \rapp modelling language consists of stereotyped class diagrams,
state machines and invariant constraints. Event handlers are indicated 
on a class diagram using the \\ {\tt <<handler>>} stereotype on a class
operation, however the body of the operation may be omitted. 
As described in figure \ref{fig:CARA-Refinement}
models are translated to Widget programs; the resulting program is a
skeleton if operation bodies are omitted from the source models. This section
specifies the translation and shows a simple example with respect to the
{\tt Main} widget and associated state machine defined in section \ref{sec:CARA-Models}.

In a \rapp model, each widget class W is associated with a function F, {\bf M}(W,F) where {\bf M} is defined as follows:
\begin{description}
\item[M0] If W is tagged {\tt <<external>>} then F is a predefined function that constructs widgets of the appropriate type.
\item[M1] If W is tagged {\tt <<widget>>} then F is a function definition with the same name. 
F has parameters specified by:
\begin{description}
\item[A1] attributes of W or of any inherited or contained widget classes (except those for {\bf A3}).
\item[A2] non-contained referenced widgets.
\item[A3] shared contained widgets (as specified by invariants).
\item[A4] the target of outgoing transitions from W and their associated arguments.
\end{description}
\noindent and a body {\bf B} that is a widget definition specified by:
\begin{description}
\item[B0] {\tt self} is used for self-reference (the default).
\item[B1] The parent of {\bf B} is a widget constructed by applying a function F' to initialization arguments. If W' is the super-class of W then {\bf M}(W',F') must hold. The initialization arguments are supplied as required by {\bf A1}--{\bf A4} in the context of F' observing any sharing constraints.
\item[B2] There is a definition in {\bf B} corresponding to each contained reference from W. If the reference is shared due to some constraint then it will have been passed as an argument. Otherwise it is constructed using the appropriate operator.
\item[B3] There is a function definition for each handler. The body of the handler must be a command. The state machine declares the target state for the handler. If this is a self-transition then the target is {\tt self}. Otherwise the transition is made by invoking the appropriate function for the target state passing any required arguments and observing any guards.
\item[B4] References to commands must be performed as command bindings.
\end{description}
\end{description}
Consider the state {\tt Main}. Applying {\bf M} produces the function shown in figure \ref{apply_M_to_Main}  (omitting type information) where the annotations on the right refer to the mapping conditions listed above. Where the rules refer to variables, they are listed in parentheses.
\begin{figure*}
\begin{center}
\begin{minipage}{0.8\textwidth}
\begin{lstlisting}
fun main(title,db,port,x,y) =                              M1, A1(title,port,x,y), A2(db)
  widget(phone(title,clock(x,y),[add(),del()]) {           M0, B0, B1
    n <- notifier(port);                                   B2
    add() = add_screen(self,db,n);                         B3, A4(self), A2(db), A3(n)
    del() = del_screen(self,db,n);                         B3, A4(self), A2(db), A3(n)
    notify(address) = do {                                 B3
      contacts <- db.records                               B4(records)
      return       
        if has_contact(addr,contacts)                      B3
        then notify() 
        else self
    };
    move(x,y) = do {                                       B3
      void <- n.move(x,y)                                  B4
      return self                                          B3(self)
    }
  }
\end{lstlisting}
\end{minipage}
\end{center}
\caption{Result of Applying {\bf M} to {\tt Main}}
\label{apply_M_to_Main}
\end{figure*}
The following section uses {\bf M} to specify function definitions for Buddy.

\subsection{Mapping Buddy}

\begin{figure*}
\begin{center}
\begin{minipage}{0.75\textwidth}
\begin{lstlisting}
type DoAdd = Widget(Button) raises add() { push:(int)-><*> raises add() }
type DoBack = Widget(Button) raises back() { push:(int)-><*> raises back() }
type DoDel = Widget(Button) raises del() { push:(int)-><*> raises del() }
type Record = { key:str; val: str }
type Main = Widget(Phone[Clock,DoAdd+DoDel]) { 
  notifier:Notifier;
  add:()-><Add>;
  del:()-><Del>;
  notify:(str)-><Notify+Main>;
  move:(int,int)-><Main>
}
type Add = Widget(Phone[AddScreen,DoAdd+DoBack]) { 
  notifier:Notifier;
  add:()-><Main>;
  back:()-><Main>;
  move:(int,int)-><Add>;
  notify:(str)-><Add>
}
type Notify = Widget(Phone[Label,DoBack]) { 
  notifier:Notifier;
  back:()-><Main>;
  move:(int,int)-><Notify>;
  notify:(str)-><Notify>
}
\end{lstlisting}
\end{minipage}
\end{center}
\caption{Type Definitions\label{fig:Type-Definitions}}
\end{figure*}

\begin{figure*}
\begin{center}
\begin{minipage}{\textwidth}
\begin{lstlisting}[numbers=left]
fun main():<Main> = do {
  contacts_db:DB[str,str] <- db[str,str]('tony_phone.dat');
  b1:<DoAdd> = widget (button('add')) { push(i:int):<*> = raise add() };
  b2:<DoDel> = widget (button('del')) { push(i:int):<*> = raise del() };
  p:Main <-
    widget self:Main (phone[Clock,DoAdd+DoDel]('Tonys Phone',clock(50,50),[b1,b2])) {
      notifier:Notifier <- create_notifier;
      add():<Add> = add_screen(self,contacts_db,notifier);
      del():<Del> = del_screen(self,contacts_db,notifier);
      notify(addr:str):<Notify + Main> = do {
        contacts:[Record] <- contacts_db.records;
        notify:Notify <- notify_screen(self,addr,notifier)
        return if has_contact(addr,contacts) then notify else self
      };
      move(x:int,y:int):<Main> = do {
        void:bool <- notifier.move(x,y)
        return self
      }
    }
  return p
}
\end{lstlisting}
\end{minipage}
\end{center}
\caption{The Main Widget\label{fig:The-Main-Widget}}
\end{figure*}

\begin{figure*}
\begin{center}
\begin{minipage}{\textwidth}
\begin{lstlisting}[numbers=left]
fun notify_screen(m:Main,addr:str,n:Notifier):<Notify> = do {
  b:DoBack = widget (button('back')) { push(i:int):<*> = raise back() };
  p:Notify <-
    widget self:Notify (phone[Label,DoBack]('Tonys Phone',label('CONTACT: ' + addr),[b])) {
      notifier:Notifier = n;
      notify(addr:str):<Notify> = do { return self };
      back():<Main> = do { return m };
      move(x:int,y:int):<Notify> = do {
        return self
      }
    }
  return p
}
\end{lstlisting}
\end{minipage}
\end{center}
\caption{The Notify Widget\label{fig:The-Notify-Widget}}
\end{figure*}

\begin{figure*}
\begin{center}
\begin{minipage}{\textwidth}
\begin{lstlisting}[numbers=left]
fun add_screen(m:Main,db:DB[str,str],n:Notifier):<Add> = do {
  records:[Record] <- db.records;
  s:<AddScreen> = addscreen(records);
  b1:<DoAdd> = widget (button('add')) { push(i:int):<*> = raise add() };
  b2:<DoBack> = widget (button('back')) { push(i:int):<*> = raise back() };
  p:Add <-
    widget self:Add (phone[AddScreen,DoAdd+DoBack]('Tonys Phone',s,[b1,b2])) {
      notifier:Notifier = n;
      notify(addr:str):<Add> = do { return self };
      add():<Main> = do { 
        name:str <- s.name;
        address:str <- s.address;
        void:str <- db.update(name,address)
        return m 
      };
      back():<Main> = do { return m };
      move(x:int,y:int):<Add> = do { return self }
    }
  return p
}
\end{lstlisting}
\end{minipage}
\end{center}
\caption{The Add Widget\label{fig:The-Add-Widget}}
\end{figure*}

The structure models in section \ref{sub:Phone-Structure} are
translated into Widget type definitions that define widget signatures
as shown in figure \ref{fig:Type-Definitions}. The type signatures
encode information about the containment structure of the widgets,
inheritance from external widgets and the state transition from section
\ref{sub:Phone-Behaviour} where each handler must return a command
that yields a widget of the appropriate type. For example, when the
\texttt{back} event in \texttt{Add} is handled, this will produce
a widget of type \texttt{Main} because of the state machine in figure
\ref{fig:State-Transitions}. The \texttt{notify} handler in \texttt{Main}
yields a widget of type \texttt{Notify} or \texttt{Main} because there
is a choice, modelled in figure \ref{fig:State-Transitions} as two
transitions with mutually exclusive guards that use \texttt{has\_contact}
to check whether an address exists in a sequence of records:

\begin{lstlisting}
rec fun has_contact(addr:str,contacts:[Record]):bool =
  if contacts = [][Record]
  then false
  else 
    let contact:Record = head[Record](contacts)
    in if contact.val = addr
       then true
       else has_contact(addr,tail[Record](contacts))
\end{lstlisting}

\noindent Each user defined root container is translated into a Widget
function that returns a command yielding a widget of the appropriate
type. Figure \ref{fig:Main-Screen} is translated into the definition
in figure \ref{fig:The-Main-Widget}. The main function returns a
command (lines 1--21) that initiates some local variables (\texttt{contacts\_db},\texttt{b1},\texttt{b2},\texttt{p})
and then yields a widget of type \texttt{Main}. The operators \texttt{clock},
\texttt{db} and \texttt{button} (lines 2,3,4,6) are built-in commands
and yield external widgets of the appropriate types. The main widget
(lines 6--19) inherits from the built-in phone widget created using
\texttt{phone} (line 6). The notifier (line 7) must perform a command
that initializes the connection to the service provider; this is done
in several steps as follows:
\begin{lstlisting}
val create_notifier:<Notifier> = do { 
  n:Notifier <- notifier(PORT); 
  void:bool <- n.connect;
  void:bool <- n.register('tony@widget.org')  
  return n 
}
\end{lstlisting}

\noindent The event handlers \texttt{add} and \texttt{del} must perform
a transition to the appropriate screen (lines 8 and 9). Notice that
in each case the transition is performed by a function that returns
a command yielding a widget of the appropriate type. The arguments
to the function allow information (\texttt{self}, \texttt{contacts\_db}
and \texttt{notifier}) to be shared between widgets.

The \texttt{notify} handler (lines 10--14) checks whether the address
of the contact that has come into range is in the receiver's contacts
database. If so then a transition to a \texttt{notify} screen is made
otherwise the command yields \texttt{self} which is a null-transition.

Finally, the \texttt{move} handler (lines 15--18) informs the notifier
of the change of location and makes a null transition.

Figure \ref{fig:The-Notify-Widget} shows the implementation of the
\texttt{Notify} widget. Notice how the use of functions allows system
states to share information by passing argument values (line 1). In
addition, since widgets can reference themselves (\texttt{self} in
figure \ref{fig:The-Main-Widget} for example) they can pass themselves
as continuation arguments (\texttt{m} in figure \ref{fig:The-Notify-Widget})
to allow the target state to make a back transition (line 9).

Figure \ref{fig:The-Add-Widget} uses the built-in \texttt{addscreen}
operation (line 3) to create a display involving the current contents
of the contacts database. An add screen supports two commands
{\tt name} and {\tt address} and is an example of a domain specific external widget
that must be realized in a platform specific way for each implementation
mapping. When the {\tt add} event is generated, the {\tt update}
command is used to change the state of the database and a transition
is made to the main screen.

\section{Related Work}\label{related_work}

As pointed out in \cite{daniele2009mda} MDA approaches have been
applied to a number of application areas, for example health care
systems \cite{jones2004formal}, however source models tend to focus
on the static structure of a system and do not include detailed behaviour.
MDA often produces code skeletons that must be edited after the PSMs
are produced. Where multiple platforms are involved, this defeats
the purpose since multiple implementations must be developed and maintained.
The approach described in \cite{daniele2009mda} uses a DSL based
on process flows defined by the A-MUSE project which differs from
the work described in this article in that widget-behaviour is expressed
using a functional programming language which is simpler, more expressive
than the A-MUSE DSL, and integrates both structural and behavioural
aspects of a system.

There are many candidates for PIM modelling languages for \rapp such as 
\cite{Sauer01uml-basedbehavior,de2006towards,reggioextension,kraemer2008engineering}.
Most of these approaches use UML class diagrams to express the structure
of an application and activity models, collaboration models and statecharts to express the behaviour. Whilst
these approaches can express any behaviour there is evidence that
behaviours can become complex \cite{cruz2010impact} and that ``{\it when a 
modeller finds two or more possible semantically equivalent 
options for modelling a system, he should pay special attention to the use of the 
constructs that are part of the components described in this work, e.g., 
reducing the total number of activities of the diagram if possible.}'' Our
view is that the use of functional abstraction in conjunction with 
state-based behaviour can significantly improve the expressiveness of
collaboration, activity, and state-machine models alone, and can form a precise foundation for many different
model driven approaches.

Mobile platforms have given rise to an interest in {\it context-aware} applications 
that react to events that arise due to changes in the state of the platform or its
environment. Context aware applications have been studied only recently and there
are few proposals for modelling notations, for example \cite{sheng2005contextuml},
where context is declared from a number of sources and triggers are
used to inform the application of context changes. Another example is \cite{tesoriero2010cauce}
where context aware applications are modelled in terms of different viewpoints: 
social, task and space and where model transformations are used to produce
platform specific models from the views.

Because of the increasing need to modularize the cross cutting context dependent behaviour, Context Oriented Programming (COP) \cite{Hirschfeld:2008fk} has been proposed. COP is a
programming paradigm to enable the expression of context-dependent
behaviour. There have been several implementations of COP languages
for Java \cite{Kamina:2010:DEC:1930021.1930023,Kamina:2011:ECP:1960275.1960305,Appeltauer:2009:IDC:1562112.1562117},
Objective-C \cite{Gonzalez:2010:SBC:1964571.1964592}, Lisp \cite{Costanza:2008:CPC:1529966.1529970},
and Smalltalk \cite{Hirschfeld:2007:ICP:1462618.1462629}. These approaches
modularize context-dependent behaviour within \emph{layers}, normally
either \emph{layer-in-class }or \emph{class-in-layer.} Layer-in-class
supports this modularization inside the class it affects. Class-in-layer
supports this modularization outside the class, being largely comparable
to aspect definitions. These languages are dependent on their
intended platform and language, leading to difficulty in porting
when attempting multi-platform development.

Given the rise of the 
target platforms (for example over 172 million
smart phones shipped worldwide in 2009 \cite{1827364}), there are
likely to be multiple DSLs or UML profiles defined for this type of
application. An MDA approach to context-aware applications is described
in \cite{almeida2006model} involving platform independent models
of different features of an application that are translated to produce
platform specific artifacts. However: how can any candidate PIM language be
evaluated and compared to others?; how can the behaviour be defined
in a universal way? 

There are a number of systems that aim to deploy a single application
across multiple mobile or web platforms. Approaches differ as described below: 
cross compilation of existing applications; a model driven approach using
a UML-style modelling language; new domain specific programming languages.

The system described in \cite{10.1109/eLmL.2010.13} has been designed
to help make code bindings between the different platform-specific
frameworks by translating Java\texttt{ .class} files to multiple platforms
including Objective-C and JavaScript. A similar approach is taken
in XMLVM \cite{Puder05,Puder06:0} where byte-code cross-compilation
is performed using a tool chain. This tool chain currently translates
Java Class files and .Net executables
 to XML documents, which then
can be output to Java byte code/.NET CIL or to JavaScript and Objective-C.
This tool chain was firstly used to cross compile Java applications
to AJAX applications \cite{1294340}, because of the lack of IDE support
and difficulty in creating an AJAX application. Further work to include
Android to iPhone application cross-compilation, as described in \cite{10.1109/eLmL.2010.13}. 

The DIMAG Framework \cite{1710058} was developed for automatic multiple
mobile platform application generation. This is accomplished by creating
a declarative definition language which is comprised of 3 distinct
parts; firstly a language DIMAG-root, which provides references to
the definitions for workflow and user interface in the application;
secondly the language State Chart eXtensible Markup Language (SCXML)
defines the workflow by the definitions of states, state transitions,
and condition based actions; and finally DIMAG-UI language based on
MyMobileWeb's IDEAL language using CSS to control the user interface.
The main shortcomings of this method is that it relies on server-side
code generation and download. 

A recent proposal for a DSL for mobile applications \cite{Behrens:2010:MID:1869542.1869562}
uses XText and Eclipse to implement a DSL that uses code generation
techniques to target mobile platforms. This DSL uses fixed GUI structures
such as \texttt{section} whereas our language uses user-declared external
widgets that integrate with the type system.
It is also not clear whether the DSL has a static type system and
its semantics is not defined independently of a translation to a target
platform.

Mobl (\texttt{\footnotesize http://www.mobl-lang.org/}) is a DSL
that has been designed to support mobile application development and
which targets JavaScript. It has many things in common with our language,
however the mobl features for describing GUI components are fixed
and the semantics is not defined independently of the target language.

Links \cite{cooper2007links} is a DSL that has been designed
to support web application development where the 3-tier architecture
is supported by a single technology. Like Widget, Links supports 
higher-order functions and is statically typed with respect to events 
and messages. Unlike our language, Links has been designed as a complete 
language with supporting tools, and indicates a possible future direction 
for layering a user language on Widget.

Web applications could not store local data to the web browser until
the development of HTML5 \cite{Wright:2009:RWO:1610252.1610260}.
In May 2007, Google released a plug-in for the Firefox web browser,
Google Gears\texttt{\footnotesize }%
\footnote{\texttt{\footnotesize http://gears.google.com/}%
}. This plug-in supports caching of web applications to allow offline
use, and also the ability for a web application to store data in a
local database. Whilst this increases the suitability of web-technology
for cross-platform \rapp systems, there are still limitations in terms
of GUI widgets a systematic use of events and static typing.

Since the arrival of HTML5 and WebKit, a number of open source and
commercial cross-platform frameworks have been proposed such as the
Appcelaterator%
\footnote{http://developer.appcelerator.com/%
}, PhoneGap%
\footnote{http://www.phonegap.com/%
} and Rhomobile%
\footnote{http://rhomobile.com/%
}. These frameworks use either JavaScript or Ruby and therefore run
in a browser. Furthermore these applications can run offline and access
the device\textquoteright{}s full capabilities; such as a GPS or camera;
providing the same look and feel as a native application.

Functional Reactive Programming (FRP) uses arrow combinators to embed
discrete event processing into the functional language Haskell \cite{hudak2004arrows,elliott2009push}.
The FRP approach is similar to the mechanism for representing commands
in Widget (in that it is monad-based) and could be used as an alternative basis for \rapp design.
However we feel that FRP is based on more abstract notions of time and events that would make the
integration described in figure \ref{fig:CARA-Refinement} more complex. Widget has been designed in terms of a domain-analysis for \rapp systems, and therefore reflects the \rapp computational framework directly in terms of event handlers, state transitions and hierarchical interfaces.

DSLs in other areas include \cite{1449858} that concentrates on the
abstraction of web applications to lower the overall complexity of
the application and boilerplate code. Further work on this DSL led
to the creation of Platform Independent Language (PIL) \cite{HemelV09}.
PIL was developed as an intermediate language, to provide a scalable
method for developing for multiple platforms. A drawback of this method
is currently it lacks support for mobile platform development. 

Other efforts for making mobile application development easier include
Google Simple%
\footnote{http://code.google.com/p/simple/%
} , a BASIC dialect for creating Android applications, and the Google
App Inventor%
\footnote{http://appinventor.googlelabs.com/about/%
} which is based on Openblocks \cite{39439454} and Kawa%
\footnote{http://www.gnu.org/software/kawa/%
}. Particularly Google App Inventor has vastly abstracted app development,
but only supports development of Android applications. These approaches are
similar to visual programming and offer a quick start for application
development but offer limited support for sophisticated behaviour.

Brenhs has proposed MDSD, a DSL for iPhone. The language is more specific
to data centric applications. Following from that work, they have
started the Applause project for developing DSL for iPhone, iPad and
Android%
\footnote{http://code.google.com/p/applause/%
}, but this is still not fully developed. 

There are a number of formal approaches to model behaviour of event-driven
systems including modal transition systems \cite{uchitel2007behaviour},
petri-nets, and the pi-calculus. Whilst these systems have good analysis
properties, they do not integrate with the structural features of
\rapp models in the way that Widget does.

In review, there are many proposals for both model-driven approaches and DSLs
for {\it rapp}. Most approaches lack an implementation independent
semantics, are unable to check system properties, many are fixed in terms of \rapp
features, and some offer limited features for expressing behaviour.

\section{Conclusion}

This article has described a rise in the interest in \rapps\ due to the
explosion of mobile, tablet and web-applications. The complexity and
proliferation of implementation technologies makes it attractive to
use model-driven techniques to develop such systems. As described
in \cite{wasserman2010software} there are a number of challenges that makes
mobile application software engineering challenging. These include 
development tools including testing, and portability. Our claim
is that \rapp models and Widget are a contribution to these challenges.
In particular, the formal definition of Widget provides scope for
tool support and analysis. A VM implementation for Widget could 
be a basis of a write once run anywhere approach 
with the associated benefits to application verification.

The Widget calculus was initially described in \cite{MobDSL} and
has been implemented as a type checker and language interpreter in
Java. The current version of the source code%
\footnote{Available from \url{http://www.eis.mdx.ac.uk/staffpages/tonyclark/Software/widget_v_1_0.zip}%
} includes the Buddy application and uses a
collection of general purpose external widgets and a phone simulator
all written in Swing. Our next step is to provide a collection of
external widgets using HTML and Android to show that the same Widget
application can run on more than one platform. In addition, we plan
to develop tooling around the \rapp modelling language that can use
the Widget calculus as a target.

\bibliographystyle{plain}
\bibliography{bib}

\end{document}